# Unconventional polarization switching mechanism in (Hf, Zr)O$_2$ ferroelectrics

Yao Wu[1,*], Yuke Zhang[1,*], Jie Jiang[1], Limei Jiang[1], Minghua Tang[1], Yichun Zhou[1],
Min Liao[1,‡], Qiong Yang[1,§], and Evgeny Y. Tsymbal[2,†]

[1] *Hunan Provincial Key Laboratory of Thin Film Materials and Devices,*
*School of Materials Science and Engineering, Xiangtan University, Xiangtan, Hunan 411105, China*

[2] *Department of Physics and Astronomy & Nebraska Center for Materials and Nanoscience,*
*University of Nebraska, Lincoln, Nebraska 68588, USA*

HfO$_2$-based ferroelectric thin films are promising for their application in ferroelectric devices. Predicting the ultimate magnitude of polarization and understanding its switching mechanism are critical to realize the optimal performance of these devices. Here, a generalized solid-state variable cell nudged elastic band (VCNEB) method is employed to predict the switching pathway associated with domain-wall motion in (Hf, Zr)O$_2$ ferroelectrics. It is found that the polarization reversal pathway, where three-fold coordinated O atoms pass across the nominal unit-cell boundaries defined by the Hf/Zr atomic planes, is energetically more favorable than the conventional pathway where the O atoms do not pass through these planes. This finding implies that the polarization orientation in the orthorhombic *Pca*2$_1$ phase of HfO$_2$ and its derivatives is opposite to that normally assumed, predicts the spontaneous polarization magnitude of about 70 μC/cm$^2$ that is nearly 50% larger than the commonly accepted value, signifies a positive intrinsic longitudinal piezoelectric coefficient, and suggests growth of ferroelectric domains, in response to an applied electric field, structurally reversed to those usually anticipated. These results provide important insights into the understanding of ferroelectricity in HfO$_2$-based ferroelectrics.

[*] These authors contributed equally to this work.
Corresponding authors: [‡]mliao@xtu.edu.cn, [§]qyang@xtu.edu.cn, [†]tsymbal@unl.edu

The modern theory of polarization predicts that the electric polarization **P** is a multi-valued quantity which is only well defined modulo "polarization quantum" $2e\mathbf{R}/\Omega$, where **R** is a lattice vector, $\Omega$ is the primitive-cell volume, and a factor of 2 stays for spin degeneracy [1-4]. Qualitatively, the multi-valued polarization is consequence of periodicity in a bulk crystal, where shifting an electron from all lattice sites by lattice vector **R** does not change the crystal structure but alters the dipole moment per unit cell by $2e\mathbf{R}$. This multi-valued polarization is believed to be irrelevant to a *change* of polarization – the quantity that can be measured in experiment. Once the initial and final polarization states are well defined, the polarization change is expected to be single valued. Nevertheless, the uncertainty remains with respect to the *direction* of ionic motion between the two oppositely polarized states. For example, in a two-dimensional ionic lattice shown in Fig. 1(a), negatively charged ions are displaced downward from their centrosymmetric positions representing polar state ①. Assuming that the lattice can be switched to a new polar state ② through centrosymmetric state Ⓒ (Fig. 1(a)) by moving the anions upward (*not* through the cation planes – N pathway), it seems to be natural to attribute polarization pointing up to the state ① and polarization pointing down to the state ②, dictated by polar displacement of anions downward or upward from their centrosymmetric positions, respectively. This picture overturns, however, if we assume that switching occurs through centrosymmetric state Ⓒ' by moving anions downward (*through* the cation planes – T pathway), as shown in Fig. 1(b), across the nominal unit cell boundary. The final state ② in Fig. 1(b) is macroscopically identical to that in Fig. 1(a), due to periodicity of the crystal structure. In this case, however, using the centrosymmetric state Ⓒ' as a reference for polar displacement of anions, it is reasonable to assign polarization pointing down to the state ① and polarization pointing up to the state ②, which is opposite to that assumed for the N pathway.

This uncertainty in the polarization direction depending on the polarization reversal pathway is accompanied by a difference in the polarization change which is expected to be the measurable quantity. Indeed, the final states ② in Figs. 1(a,b) are related by one unit cell translation of the anion sublattice, making these states macroscopically identical but belonging to two separate polarization branches differed by polarization quantum $2ec/\Omega$, where $c$ is the lattice constant in the upright direction of the lattice. As a result, polarization change $\Delta P_N$ for the N pathway is related to polarization change $\Delta P_T$ for the T pathway as $\Delta P_T = \Delta P_N + 2ec/\Omega$. Thus, depending on the polarization switching pathway, the polarization change (a measurable quantity) has opposite sign and is different by the polarization quantum.

This observation has implications for the domain-wall motion in response to electric field. For the N pathway of switching in Fig. 1(a), the electric field pointing down pushes negative ions upward resulting in growth of domain ② in expense of domain ① (Fig. 1(c), top). In this case, the domain wall (DW) moves from right to left. If the electric field is reversed, the switching process is also reversed, and the DW moves from left to right (Fig. 1(c), bottom). This behavior overturns for the T pathway (Fig. 1(b)). In this case, the electric field pointing down forces domain ① to grow while domain ② to shrink, resulting in the DW motion from left to right (Fig. 1(d), top). For the electric field pointing up, the switching process is reversed and the DW moves from right to left (Fig. 1(d), bottom). We see therefore that depending on the polarization switching pathway, DW motion occurs in opposite directions.

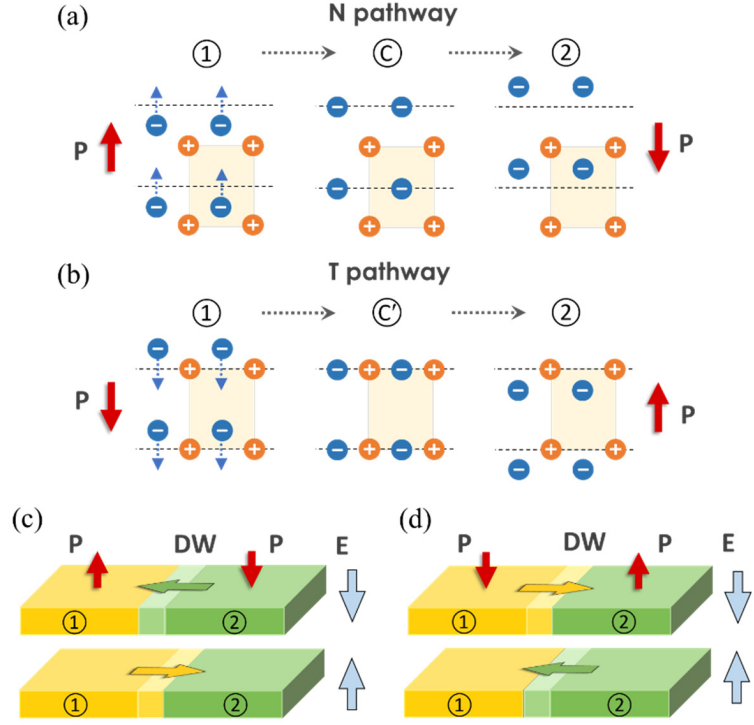

FIG.1 (a,b) Polarization reversal in a two-dimensional ionic lattice from state ① to state ②, where negatively charged ions move upward (N pathway) (a) or downward (T pathway) (b) crossing their centrosymmetric positions (indicated by the dashed lines) in states Ⓒ and Ⓒ', respectively. Red arrows and blue dashed arrows indicate polarization orientations and directions of ionic motion, respectively. (c,d) Polarization switching process in response to applied electric field occurring through domain-wall motion via N (c) and T (d) pathways. Polarization (P) and electric field (E) directions are indicated by red and blue arrows, respectively. Horizontal green and yellow arrows indicate directions of domain-wall motion.

    The polarization dependence on the switching pathway is *not* essential for most ferroelectric materials due to displacement of ions from their centrosymmetric positions being very small compared to the lattice constant. For example, in a perovskite ferroelectric BaTiO$_3$, polar displacement of Ti atoms is ~0.1Å, while the lattice constant is ~4Å. In this case, polarization reversal occurs through a well-defined centrosymmetric phase (*Pm*-3*m* for BaTiO$_3$), while the ionic motion across the unit cell boundaries is energetically prohibitive.

    The situation is however different for ferroelectric HfO$_2$. This recently discovered ferroelectric material has advantages of being compatible with the conventional complementary metal-oxide-semiconductor (CMOS) technology, having robust ferroelectricity at the nanometer scale, and thus being promising for device applications [5-13]. Up to date, several crystallographic phases, such as orthorhombic *Pca*2$_1$ (*Pbc*2$_1$) [5,14-17] and *Pmn*2$_1$ [18,19] and rhombohedral *R*3*m* [20,21], have been considered to support ferroelectricity in HfO$_2$. Very recently, Yun *et al*. [22] have unambiguously associated *intrinsic* ferroelectricity in epitaxially grown HfO$_2$ films with the orthorhombic *Pca*2$_1$ phase. While there are still debates regarding its stabilization mechanism [23-26], this structural phase is known to represent a lateral array of vertically aligned polar columns of HfO$_2$ separated by nonpolar columns (Fig. 2(a)). The vertical displacement of the three-fold

coordinated oxygen atoms from their centrosymmetric positions in the polar columns is about 0.56Å. This displacement is comparable to that, 0.71Å, evaluated with respect to the unit cell boundary, which makes both the N and T switching pathways possible (Fig. 2), resulting in the ambiguity of the polarization direction, as well as its magnitude. Thus, understanding the microscopic mechanism of polarization switching in HfO$_2$ is not only important *per se*, but also critical for predicting the polarization magnitude that is measured in experiment.

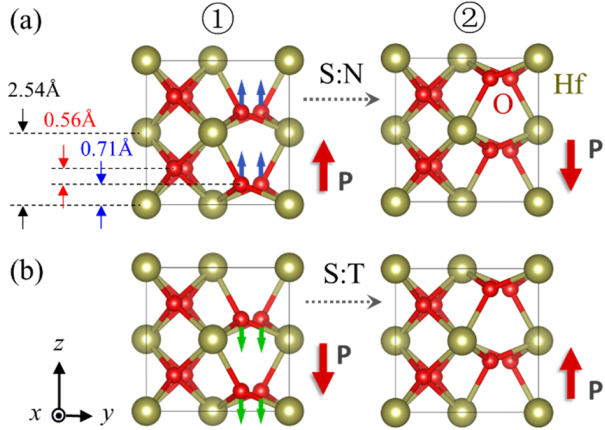

FIG.2 Atomic structure of HfO$_2$ and polarization switching pathways between states ① and ②. S:N (a) and S:T (b) pathways where three-fold coordinated O atoms not pass through and passing through the Hf atomic planes, respectively. O atom displacements are indicated by blue and green arrows. Red arrows show pathway-dependent polarization directions for states ① and ②.

Previous studies have largely focused on the 180° uniform polarization switching in HfO$_2$-based ferroelectrics where the whole uniformly polarized domain reversed its polarization [18,27-29]. It was argued that possible pathways of polarization switching can be divided into two main categories depending on the motion direction of three-fold coordinated O atoms – passing through (T pathway) or not passing through (N pathway) the Hf atomic planes. It was found that the T pathway (going across the orthorhombic *Pbcm* phase) has a much higher energy barrier than the N pathway (going across the tetragonal *P*4$_2$/*nmc* phase) (Table SII in Supplemental Material [30] summarizing these results). It is known however that polarization switching in ferroelectrics is realized via nucleation and growth of ferroelectric domains of reversed polarization, rather than uniform polarization switching [31]. Based on the *most stable* 180° DW structure, the energy barrier for DW motion, where polarization switching occurs via the N pathway, was calculated to be about 1eV/u.c. [32,33]. (A much lower switching energy barrier was predicted for a topological DW [34], but such a DW has ~1eV/u.c. higher formation energy.) However, no studies have been performed for polarization switching associated with DW motion via the T pathway.

To address this deficiency, in this letter, we employ a generalized solid-state variable cell nudged elastic band (VCNEB) method [35] to predict the most energetically favorable switching pathway in ferroelectric HfO$_2$ and its derivatives. Using this approach, we consistently study polarization reversal pathways in (Hf,Zr)O$_2$ ferroelectrics associated with the DW motion and compare them with other switching mechanisms. We demonstrate that the pathway where O atoms pass through the Hf/Zr atomic planes has the lowest potential barrier thus challenging the previously found results. This finding has significant implications for the understanding of the polarization reversal mechanism in HfO$_2$-based ferroelectrics, the assignment of polarization orientation to different ferroelectric domains, DW motion under an applied electric field, and the measured magnitude of ferroelectric polarization and piezoelectric response in these materials.

We consider 180° DW motion as the primary mechanism for polarization switching, whereas uniform polarization switching is taken as a reference (see Supplemental Material [30] for computational details). As follows from the previous theoretical studies [32,33] and atomic-scale characterization [36], the most favorable 180° DW structure in HfO$_2$-based ferroelectrics represents an atomically sharp interface between domains ① and ② mimicking the *Pbca* phase. While uniform polarization reversal may involve displacement of three-fold coordinated O atoms either (nearly) *straight* or *crosswise* perpendicular to the DW [30], the only straight O displacement is permitted in the process of 180° DW motion. These pathways of polarization reversal are denoted in Fig. 2 by S:N and S:T and are analogous to the N and T pathways in Fig.1.

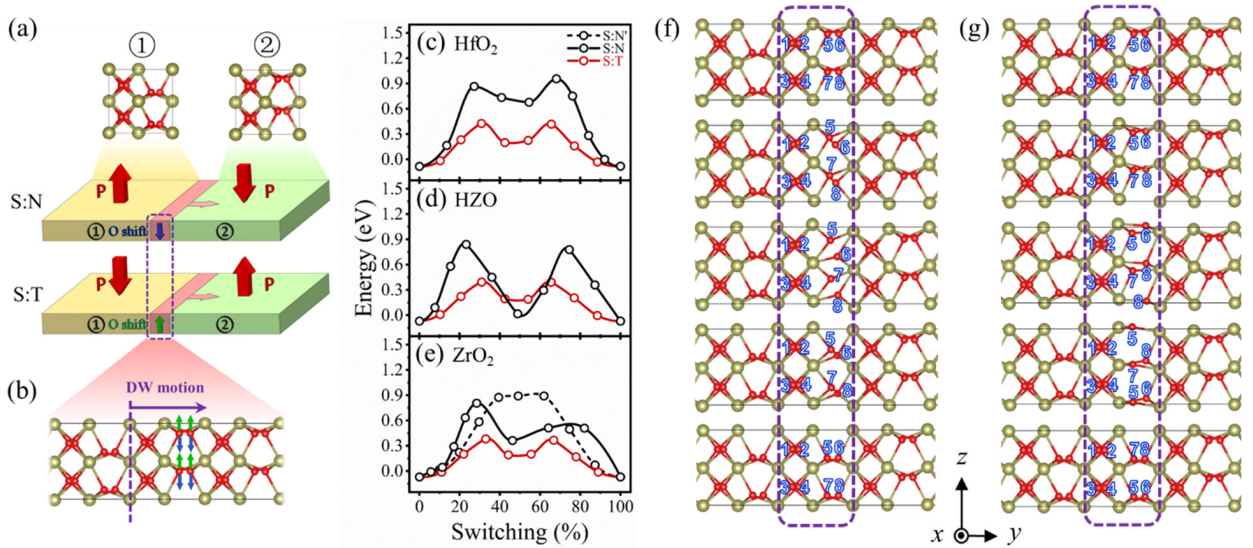

FIG.3 (a) Sketch of 180° DW motion between domains ① and ② along the S:N and S:T pathways in (Hf,Zr)O$_2$ ferroelectrics. Red arrows indicate polarizations directions of the two ferroelectric domains. (b) DW atomic structure with indicated shifts of three-fold coordinated O atoms for the DW motion along S:N (blue arrows) and S:T (green arrows) pathways. (c-e) Energy profiles for DW motion in HfO$_2$ (c), HZO (d), and ZrO$_2$ (e). The total energy of a uniformly polarized domain is set to zero. (f,g) Atomic evolution of DW motion in HfO$_2$ along S:N (f) and S:T (g) pathways. Dashed rectangles indicate the unit cell where polarization reversal occurs.

Fig. 3(a) sketches 180° DW propagation along the S:N and S:T pathways. The DW is set to move rightward, as shown in Fig. 3(b), driven by displacements of three-fold coordinated O atoms along the S:N and S:T pathways (blue and green arrows in Fig. 3(b)) resulting in polarization reversal within the unit cell at the DW. Figs. 3(c-e) (black solid lines) show the calculated energy profiles for DW propagation along the S:N pathway in HfO$_2$, HZO, and ZrO$_2$. DW motion exhibits double-barrier energy profiles with the barrier heights of about 1eV (Table I). Fig. 3(f) shows the structural evolution in the process of DW motion for HfO$_2$ (Figs. S9 and S10 for HZO and ZrO$_2$).

DW displacement by one unit cell occurs via a successive downward shift of three-fold coordinated O atoms in a unit cell. This leads to an intermediate state, resembling a monoclinic $P2_1/c$ phase (distorted tetragonal $P4_2/nmc$ phase), which is responsible for the local minimum in the energy profiles in Figs. 3(c-e). Due to different local symmetry conditions, the structural evolution along the S:N pathway for the DW motion is different from that obtained for uniform polarization reversal. While, for uniform switching, O atoms shift toward the unit cell center leading to an intermediate state of the tetragonal $P4_2/nmc$ phase of low energy (Fig. S2(a)), for DW motion, this process is prohibited to meet structural continuity with the neighboring cells (Fig. 3(f)) leading to a higher energy barrier (Table I). The simultaneous downward displacement mode of the S:N pathway (named the S:N' pathway and characterized by structural evolution shown in Fig. S11) is only stable in $ZrO_2$ (black dashed line in Fig. 3(e)).

Polarization switching along the S:T pathway reveals a different behavior. While the calculated energy curves for all three ferroelectrics $HfO_2$, HZO, and $ZrO_2$ also exhibit two peaks along the S:T pathway (Figs. 3(c-e), red solid lines), the associated energy barriers are only about 0.5 eV, nearly half the barriers for the S:N pathway (Table I). As seen from Fig. 3(g) for $HfO_2$ (Figs. S12 and S13 for HZO and $ZrO_2$), upward movement of the O atoms through the Hf atomic planes goes through an $P2_1/c$ intermediate state (distorted $Pbcm$ phase), producing the energy minimum. This structural evolution is qualitatively similar to that found for the uniform S:T switching of $HfO_2$ (Fig. S2(b), bottom path) and leads to the comparable energy barriers (Table I).

The significantly lower barrier height for the S:T pathway compared to the S:N pathway can be understood from the structural evolution displayed in Fig. S14. The top (yellow arrows) and bottom (cyan arrows) three-fold coordinated O atoms have opposite displacements along the $x$-direction that are reversed in one domain with respect to the other. When the DW moves along the S:N pathway (Fig. S14(a)), the polar shift of the O atoms along the $-z$-direction is accompanied by their crosswise displacement along the $x$-direction. This leads to structural discontinuity at the DW in the intermediate state and results in a considerable energy cost. On the contrary, when the DW moves along the S:T pathway (Fig. S14(b)), the top three-fold coordinated O atoms in one domain continuously transform into the bottom ones in the other domain due to the polar shift of the O atoms along the $+z$-direction with no crosswise displacement in the $x$-direction. This makes the intermediate state for the S:T pathway closer to the ground-state monoclinic $P2_1/c$ phase resulting in the substantially reduced energy barrier.

TABLE I. Polarization ($P$), polarization quantum ($2ec/\Omega$), and energy barrier per unit cell ($E_b$) for uniform polarization (UP) reversal, domain-wall (DW) assisted switching, and unit cell (UC) switching in a single domain for $HfO_2$, HZO, and $ZrO_2$ calculated using the VCNEB method for S:N and S:T pathways.

| Crystal | $HfO_2$ | | HZO | | $ZrO_2$ | |
|---|---|---|---|---|---|---|
| Pathway | S:N | S:T | S:N | S:T | S:N | S:T |
| $P$ (μC/cm$^2$) | 51.0 | 69.7 | 51.9 | 67.3 | 52.3 | 65.5 |
| $2ec/\Omega$ (μC/cm$^2$) | 120.8 | | 119.2 | | 117.7 | |
| $E_b$ (eV): UP | 0.461 | 0.432 | 0.349 | 0.377 | 0.293 | 0.362 |
| $E_b$ (eV): DW | 1.040 | 0.509 | 0.911 | 0.464 | 0.876 | 0.452 |
| $E_b$ (eV): UC | 0.918 | 0.517 | 0.860 | 0.480 | 0.786 | 0.427 |

Importantly, the T pathway appears to be the most favorable not only for the DW motion but also for nucleation of a domain with reversed polarization. We simulate this process by the S:N and S:T polarization reversal within a unit cell of $(Hf,Zr)O_2$ single-domain ferroelectrics [30]. It appears that the S:T pathway exhibits about half the energy barrier of the S:N pathway independent of $(Hf,Zr)O_2$ stoichiometry (Table I). Thus, both domain nucleation and their growth can efficiently occur through the S:T switching mechanism. We note here that the similar energy barriers for unit-cell switching at the DW and within a uniformly polarized domain indicate that the polarization reversal process in $HfO_2$ may be controlled both by domain nucleation and growth and by random unit-cell switching, as has been observed experimentally [37].

Our results challenge the conventional understanding of polarization switching in $HfO_2$-based ferroelectrics and have important implications. *First*, the S:T switching pathway being responsible for polarization reversal implies that polarization pointing down needs to be assigned to the domain ① while the polarization pointing up to the domain ②, as shown in Fig. 2(b). This assignment is at odds with the conventional picture where the displacement of the three-fold coordinated O ions from their centrosymmetric positions in the tetragonal $P4_2/nmc$ phase is regarded to be polar leading to the polarization orientations shown in Fig. 2(a). It appears that the orthorhombic *Pbcm* phase (or its distorted $P2_1/c$ variant) needs to be considered as the centrosymmetric phase reference, leading to polarization orientation shown in Fig. 2(b). The dependence of polarization sign on the switching pathway has been also pointed out recently in Refs.[34,38].

*Second*, the ultimate polarization magnitude that can be measured in experiment appears to be different from that conventionally assumed. Polarization is calculated using the standard Berry phase method [2] and exhibits several branches separated by the polarization quantum $2ec/\Omega$ (Fig. S24). The S:N and S:T pathways reveal opposite slopes spanning the total polarization range of $2\times 2ec/\Omega$. This implies that the polarization change has different signs for the S:N and S:T pathways and their absolute values, $P_N$ and $P_T$, add up to the value of $2ec/\Omega$ (≈120.8 μC/cm² for $HfO_2$). Table I shows polarization values calculated for the S:N and S:T pathways, indicating that within the computational accuracy their sum is $2ec/\Omega$ for all three $(Hf,Zr)O_2$ stoichiometries considered. Importantly, the predicted $P_T$ value of about 70μC/cm² corresponding to the most energetically favorable switching pathway is larger than the $P_N$ value of about 50μC/cm² that is commonly anticipated for $HfO_2$. This result is plausible for potential application of $HfO_2$-based ferroelectrics where a higher polarization implies a stronger response to external stimulus.

We note that there is a lot of controversy in the literature regarding the experimentally measured values of polarization. This is due to fluctuating quality of films grown in different laboratories, effects of grain boundaries, and defects such as oxygen vacancies [39] which are detrimental to the intrinsic ferroelectricity of $HfO_2$. Recently, however, Yun *at al*. [22] were able to grow Y-doped $HfO_2$ films with a high degree of crystallinity exhibiting a ferroelectric response free from the ambiguities associated with oxygen vacancies and grain boundaries. The measured *intrinsic* polarization was found to be 64 μC/cm², i.e., much larger than the nominal 50 μC/cm², indirectly signaling the polarization switching mechanism predicted in this work.

The *third* implication following from our results is the DW motion in response to electric field. Applying an electric field up in Fig. 2 is expected to stimulate growth of domain ② with polarization parallel to the applied field (Fig. 2(b)), rather than domain ① (Fig. 2(a)) anticipated in the conventional picture. This prediction can be verified experimentally using the recently developed *in-situ* biasing technique in scanning transmission electron microscopy (STEM) [40]. While the precise detection of O atom displacements is challenging, recent advances in STEM make it feasible (e.g., Refs. [15,39]).

The *fourth* implication is related to the longitudinal piezoelectric coefficient $d_{33}$. Experimentally, $d_{33}$ varies in magnitude and sign depending on film thickness, deposition method, sample history, *etc*. [41]. Theoretically, for the conventional polarization direction (Fig. 2(a)), $d_{33}$ is predicted to be *negative* [42,43] due to the preserved equilibrium distance of the Hf-O bonds in the switching process. For the opposite polarization associated with the S:T switching pathway (Fig. 2(b)), this mechanism leads to a *positive* intrinsic longitudinal piezoelectric coefficient. Different sign of the piezoelectric response depending on the polarization switching pathway has been discussed recently by Qi *et al*. [38]. The predicted unconventional switching may also impact pyroelectricity of $HfO_2$ [44].

In summary, we have predicted that ferroelectric polarization switching in $(Hf,Zr)O_2$ ferroelectrics associated with domain-wall motion occurs through the S:T pathway, where three-fold coordinated O atoms pass across the nominal unit-cell boundaries defined by the Hf/Zr atomic planes, rather than the conventional S:N pathway. This finding implies that the polarization orientation in the orthorhombic $Pca2_1$ phase of $HfO_2$ and its derivatives is opposite to that normally assumed, predicts the spontaneous polarization magnitude of about $70\mu C/cm^2$ that is nearly 50% larger than the commonly accepted value, signifies a positive intrinsic longitudinal piezoelectric coefficient, and suggests growth of ferroelectric domains, in response to an applied electric field, structurally reversed to those usually anticipated. Our predictions are important for the understanding of polarization behavior in $HfO_2$-based ferroelectrics, and therefore we hope that they will stimulate efforts to verify them experimentally.

The authors thank Alexei Gruverman and Xiaoshan Xu for useful discussions. This work was supported by the National Natural Science Foundation of China (Grant Nos. 12072307, 52072324, and 92164108), the Outstanding Youth Science Foundation of Hunan Province, China (Grant No. 2021JJ20041), and the Research Foundation of Education Bureau of Hunan Province, China (Grant Nos. 21A0114, 21B0112, and 191A06). E.Y.T. acknowledges support from the National Natural Science Foundation through the EPSCoR RII Track-1 program (NSF Award OIA-2044049). The atomic structures were plotted using VESTA software.

# SUPPLEMENTAL MATERIAL

# Unconventional polarization switching mechanism in (Hf, Zr)O$_2$ ferroelectrics

Yao Wu[1,*], Yuke Zhang[1,*], Jie Jiang[1], Limei Jiang[1], Minghua Tang[1], Yichun Zhou[1], Min Liao[1,‡], Qiong Yang[1,§], and Evgeny Y. Tsymbal[2,†]

[1] *Hunan Provincial Key Laboratory of Thin Film Materials and Devices, School of Materials Science and Engineering, Xiangtan University, Xiangtan, Hunan 411105, China*

[2] *Department of Physics and Astronomy & Nebraska Center for Materials and Nanoscience, University of Nebraska, Lincoln, Nebraska 68588, USA*

**I. Computational methods**

Density functional theory (DFT) calculations are performed within the Vienna *ab-initio* simulation package (VASP) by using the projector-augmented wave (PAW) method [1]. The Perdew-Burke-Ernzerhof generalized gradient-approximation (GGA-PBE) [2] is adopted to treat the exchange-correlation interaction. The widely accepted orthorhombic phase *Pca*2$_1$ [3-7] is taken as the ferroelectric phase of (Hf, Zr)O$_2$. In this structure (Fig. 2 in the main text), Hf atoms are located at the corners and face centers of the orthorhombic unit cell which contains 12 atoms. The left column of the unit cell contains four-fold coordinated O atoms located near the center of Hf tetrahedrons, while the right column contains three-fold coordinated O atoms which are displaced from the center of Hf tetrahedrons and responsible for ferroelectric polarization. The cutoff energy of 500 eV and a 7 × 7 × 7 Monkhorst Pack k-mesh are used in the calculations. The atomic positions and lattice constants are fully relaxed until the atomic Feynman forces are smaller than 0.005 eV/Å. Hf$_{0.5}$Zr$_{0.5}$O$_2$ (HZO) is modelled assuming that Zr atoms homogeneously substitute Hf atoms in HfO$_2$. This approach is justified based on the results of the recent study [8] showing that the phase stability of Zr-doped HfO$_2$ is independent of the Zr atom ordering. The calculated lattice parameters of bulk *Pca*2$_1$ HfO$_2$ are $a$ = 5.258 Å, $b$ = 5.046 Å, and $c$ = 5.071 Å. Lattice parameters of HZO are $a$ = 5.292 Å, $b$ = 5.080 Å, and $c$ = 5.104 Å. Lattices parameters of ZrO$_2$ are $a$ = 5.322 Å, $b$ = 5.117 Å, and $c$ = 5.137 Å.

Both the generalized solid-state nudged elastic band method (with variable-cell, abbreviated VCNEB) [9] and conventional climbing image nudged elastic band method (with non-variable cell, abbreviated nVCNEB) [10] are used to simulate polarization reversals in *Pca*2$_1$ HfO$_2$, HZO, and ZrO$_2$. Four polarization reversal pathways sketched in Figs. S1(a,b) are considered for uniform polarization switching. Since the three-fold and four-fold coordinated O atomic columns are always alternatively arranged normal the domain-wall direction [11], polarization reversal with exchange of three-fold and four-fold coordinated O atoms (e.g., pathway (2) in Fig. S1 of Ref. [12]) is not taken into consideration due to an obviously high energy cost for this process. The minimum energy paths for polarization reversal are obtained until the forces on each atom are converged to become smaller than 0.02 eV/Å in both VCNEB and nVCNEB simulations. The Berry phase method [13] is used to calculate the spontaneous polarization for the four uniform polarization reversal pathways.

The domain wall (DW) structure is comprised of the ① and ② variants shown in Fig. 2 in the main text and has a negative DW energy. This 180° DW structure possesses continuity of O atomic displacements perpendicular to the DW plane (along the $y$-direction). To simulate the rightward shift of the DW across one unit cell, the initial state of the DW structure is assumed to contain 3 unit cells of variant ① and 5 unit cells of variant ②, and the final state contains 4 unit cells of variant ① and 4 unit cells of variant ②. The DW motion across one unit cell is studied by simulating the polarization reversal of a unit cell at the DW (marked by dashed rectangles in Figs. 3(f, g)) similar to our previous studies [6,14]. A $5 \times 1 \times 5$ Monkhorst Pack k-mesh is used in the calculation of the DW motion. Nucleation of a one-unit-cell domain with reversed polarization is simulated for comparison.

**II. Uniform polarization reversal**

As shown in Figs. S1(a) and (b), 180° uniform polarization reversal can transform initial state structure ① into one of the two final state structures, ② or ③ [6]. Transition from the structure ① to the structure ② occurs via almost straight motion of the three-fold coordinated O atoms either upward by not passing through the Hf atomic planes, named S:N pathway (Straight: Not through) or downward by passing through the Hf atomic planes, named S:T pathway (Straight: Through). In contrast, transition from the structure ① to the structure ③, occurs via crosswise along the $y$-direction motion of the three-fold coordinated O atoms either upward by not passing through the Hf atomic planes, named C:N pathway (Crosswise: Not through) or downward by passing through the Hf atomic planes, named C:T pathway (Crosswise : Through).

First, polarization reversal from state ① to state ② along the S:N pathway is studied for $HfO_2$. The black curve in Fig. S1(c) shows the corresponding energy profile. The energy barrier of polarization switching for this pathway is 0.461 eV (Table SI). The structural evolution along this pathway shown in Fig. S2(a) indicates that the intermediate state is close to the tetragonal $P4_2/nmc$ phase. The energy barrier is much lower than that calculated within the nVCNEB method 0.695 eV (Fig. S1(d) and Table SII), which is consistent with our previous calculations [6] and other reported results [5,15,16].

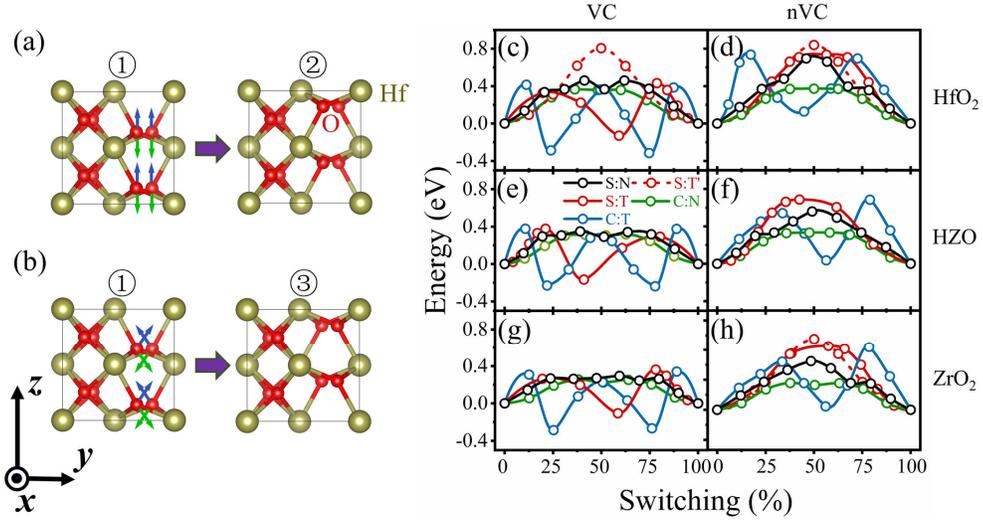

FIG. S1. (a, b) Uniform polarization reversal pathways of $HfO_2$ ferroelectric in $Pca2_1$ phase. Blue and green arrows represent the directions of motion of the three-fold coordinated O atoms during polarization reversal. (c-h) The energy profiles of uniform polarization reversal pathways for (c, d) $HfO_2$, (e, f) HZO, and (g, h) $ZrO_2$ per unit cell calculated within the (c, e, g) VCNEB and (d, f, h) nVCNEB methods. The energy of the $Pca2_1$ unit cells (initial and final states) is set to be zero.

TABLE SI. The energy barriers calculated with the VCNEB method ($E_b$, per unit cell) and spontaneous polarizations ($P$) of the four uniform polarization reversal pathways for $HfO_2$, HZO, and $ZrO_2$. Previous published results for $HfO_2$ are shown for comparison.

| Path-way | $HfO_2$ | | HZO | | $ZrO_2$ | | $HfO_2$ (previous work) | |
|---|---|---|---|---|---|---|---|---|
| | $E_b$ (eV) | $P$ ($\mu C/cm^2$) | $E_b$ (eV) | $P$ ($\mu C/cm^2$) | $E_b$ (eV) | $P$ ($\mu C/cm^2$) | $E_b$ (eV) | $P$ ($\mu C/cm^2$) |
| S:N | 0.461 | 51.0 | 0.349 | 51.9 | 0.293 | 52.3 | 0.440[16] | 50[5], 51.8[6], 52[17,18] |
| S:T' | 0.805 | 69.7 | / | 67.3 | / | 65.5 | / | 66[19], 66.3[20], 68[21], 69[17], 70[5] |
| S:T | 0.432 | | 0.377 | | 0.362 | | / | |
| C:N | 0.366 | 50.9 | 0.317 | 51.6 | 0.254 | 51.7 | 0.480[18], 0.288[16] | 50[5,22], 51.8[6], 52[17,18] |
| C:T | 0.415 | 69.9 | 0.376 | 67.6 | 0.340 | 66.0 | / | 68[21], 69[17] |

TABLE SII. The energy barriers ($E_b$, per unit cell, in eV) of the four uniform polarization reversal pathways for HfO$_2$, HZO, and ZrO$_2$ calculated within nVCNEB method. Previous published results for the energy barriers of HfO$_2$ are listed for comparison.

| Pathway | HfO$_2$ | HZO | ZrO$_2$ | HfO$_2$ (previous work) |
|---|---|---|---|---|
| S:N | 0.695 | 0.559 | 0.486 | 0.712[5], 0.63[6], 0.65[15], 0.66[17], 0.688[16] |
| S:T' | 0.839 | / | 0.713 | 0.8[5,22], 0.84[19,21], 0.788[17], 0.929[20] |
| S:T | 0.736 | 0.685 | 0.640 | / |
| C:N | 0.373 | 0.334 | 0.267 | 0.336[5], 0.4[3,23], 0.36[21], 0.35[6], 0.368[15], 0.372[24], 0.3375[17], 0.316[16] |
| C:T | 0.736 | 0.684 | 0.629 | 0.831[15], 0.705[17] |

Then, we examine the polarization reversal from state ① to state ② along the S:T pathway in HfO$_2$. It is interesting that the two modes of structural evolution with different energy barriers seem feasible for this pathway. The corresponding energy profile of the first structural evolution mode (named S:T' pathway) is shown by the red dashed curve in Fig. S1(c). It is seen that the S:T' pathway exhibits a single-barrier potential energy curve with the highest barrier height of 0.805 eV among the four pathways. This barrier height is comparable to that (0.839 eV) calculated within the nVCNEB method (Fig. S1(d), red dashed curve, and Table SII). From the structural evolution along this pathway shown in the top panels of Figs. S2(b) and S3(b), it is seen that the intermediate state of HfO$_2$ is the orthorhombic *Pbcm* phase, resulting from the simultaneous downward shift of the three-fold coordinated O atoms, which is in agreement with the previous results [5,15,25].

The corresponding energy profile of the second structural evolution mode (named S:T pathway) is shown by the red solid curve in Fig. S1(c). A deep potential well appears in the energy curve for this pathway, which substantially lowers the polarization reversal energy barrier to 0.432 eV. The structural evolution for the S:T pathway (Fig. S2(b), bottom panel) indicates that the three-fold coordinated O atoms shift downward successively during polarization reversal, leading to an intermediate state of the monoclinic *P*2$_1$/*c* phase, which is more stable than the initial and final *Pca*2$_1$ states. This S:T pathway with the successive three-fold coordinated O atoms shift also appears in the nVCNEB simulation (Fig. S3(b), bottom panel). However, as follows from the red solid curve in Fig. S1(d) and Table SII, the barrier height is still 0.736 eV, and the potential well doesn't appear due to the applied strain constraint. This S:T pathway with the successive three-fold coordinated O atoms shift has never been reported in the previous studies. However, this S:T pathway seems more favorable based on its much lower energy barrier height as compared to the S:T' pathway, if the lattice deformation during the polarization reversal is tolerated.

Next, the polarization reversal from state ① to state ③ along the C:N or C:T pathways is explored in HfO$_2$. As seen from the green curve in Fig. S1(c) and Table SI, the C:N pathway has the energy barrier of 0.366 eV, similar to that (0.373 eV) calculated within the nVCNEB method (Fig. S1(d), green curve, and Table SII). From the structural evolution shown in Figs. S2(c) and S3(c), it is seen that the polarization reversal pathway goes through an intermediate state close to the tetragonal $P4_2/nmc$ phase. The blue curve in Fig. S1(c) shows the potential energy profile for the polarization reversal of $HfO_2$ along the C:T pathway. It is surprising that the energy profile shows two potential wells with the energy barrier of 0.415 eV. The structural evolution for this pathway (Fig. S2(d)) is more complicated in comparison to the other three pathways. The successive downward shift of the three-fold coordinated O atoms and the accompanied crosswise displacement of O atoms along the $y$-direction lead to two monoclinic $P2_1/c$ phases with opposite lattice angles at the two potential wells. The saddle point between the two potential wells is the $P4_2/nmc$ phase. In contrast, the energy profile for the C:T pathway calculated within the nVCNEB method shows two peaks (Fig. S1(d), blue curve) with 0.736 eV barrier height caused by the prior downward shift of the three-fold coordinated O atoms and the followed crosswise displacement of the O atoms along the $y$-direction (Fig. S3(d)). The first half of the C:T pathway within the nVCNEB method appears to be similar to the S:T pathway.

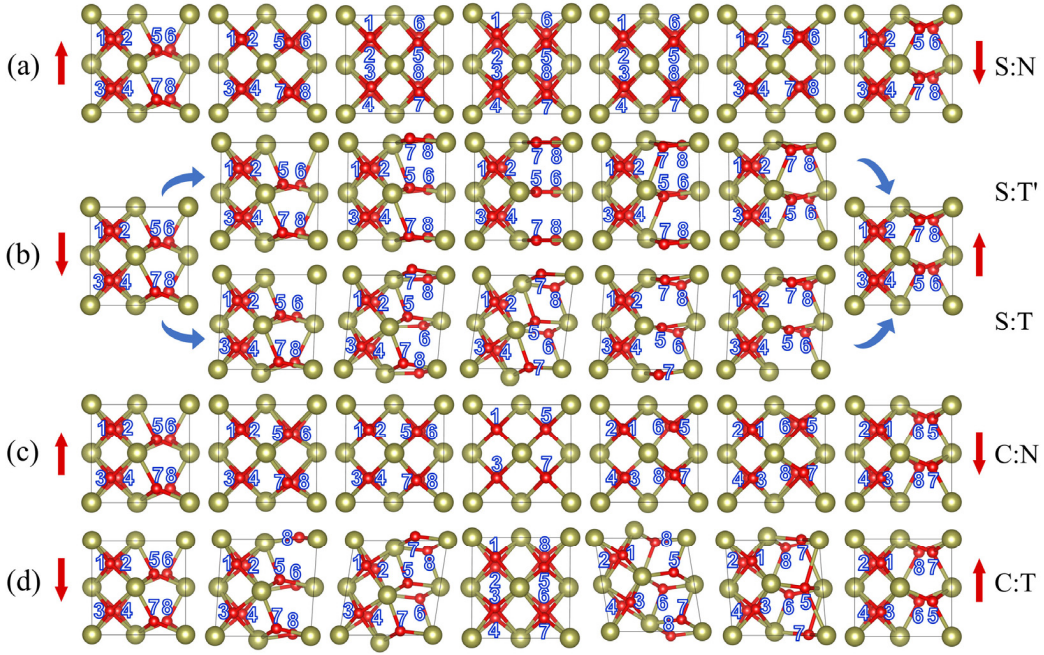

FIG. S2. The structural evolution of the four uniform polarization reversal pathways of $HfO_2$ calculated within the VCNEB method. (a) S:N pathway. (b) S:T' (top panel) and S:T (bottom panel) pathways. (c) C:N pathway. (d) C:T pathway. Red arrows represent the ferroelectric polarization directions of the initial and final states. Numbers on the atomic structures track the motion of O atoms.

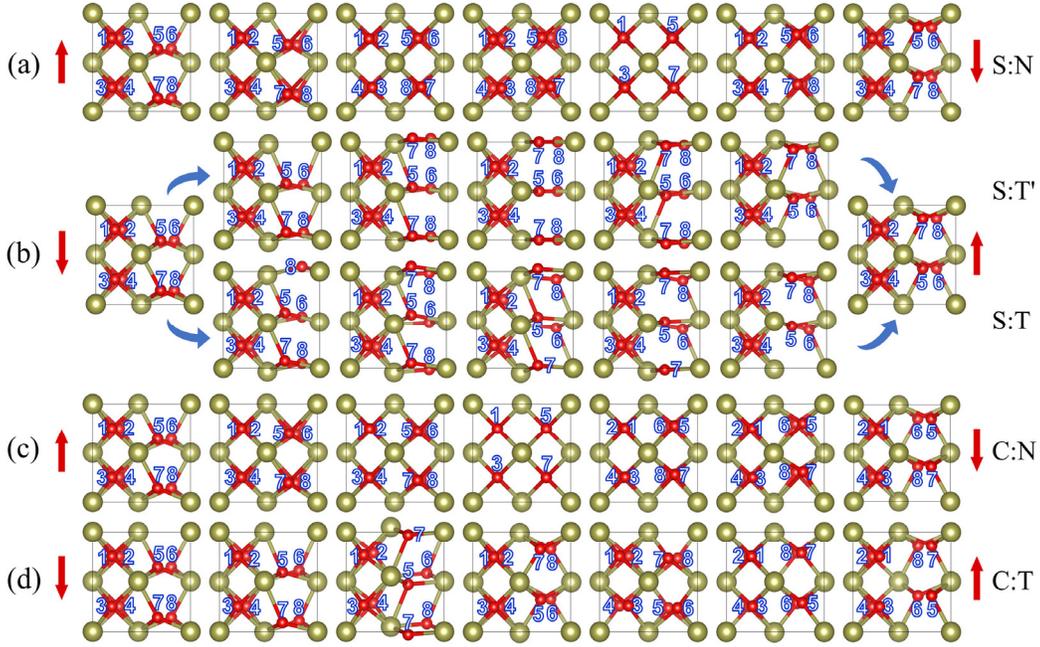

FIG. S3. The structural evolution of four uniform polarization reversal pathways of $HfO_2$ calculated within the nVCNEB method. (a) S:N pathway. (b) S:T' (top panel) and S:T (bottom panel) pathways. (c) C:N pathway. (d) C:T pathway.

Then, the effect of Zr doping on the uniform polarization reversal is explored. It is found that $Hf_{0.5}Zr_{0.5}O_2$ (HZO) and $ZrO_2$ exhibit similar structural evolution behaviors (Figs. S4-S7) to those of $HfO_2$. However, the energy barriers of all the polarization reversal pathways gradually decrease when the Hf sites of $HfO_2$ are partially (HZO) or completely ($ZrO_2$) substituted by Zr (Figs. S1(e-h), Tables SI and SII). Furthermore, the S:T' pathway is not stable in HZO and $ZrO_2$, which may be due to the doping-induced symmetry reduction. As follows from the comparison between the four pathways, the C:N pathway for (Hf, Zr)$O_2$ has the lowest polarization reversal energy barrier within both the VCNEB and nVCNEB calculations, which is also in agreement with the previous results shown in Table SII [6,15,18,23]. However, the barrier height differences between the four pathways calculated within the VCNEB method are not as significant as those within the nVCNEB method (Figs. S1(c-h)), because the dynamic lattice deformation of the S:N, S:T, and C:T pathways are tolerated within the VCNEB method (Fig. S8).

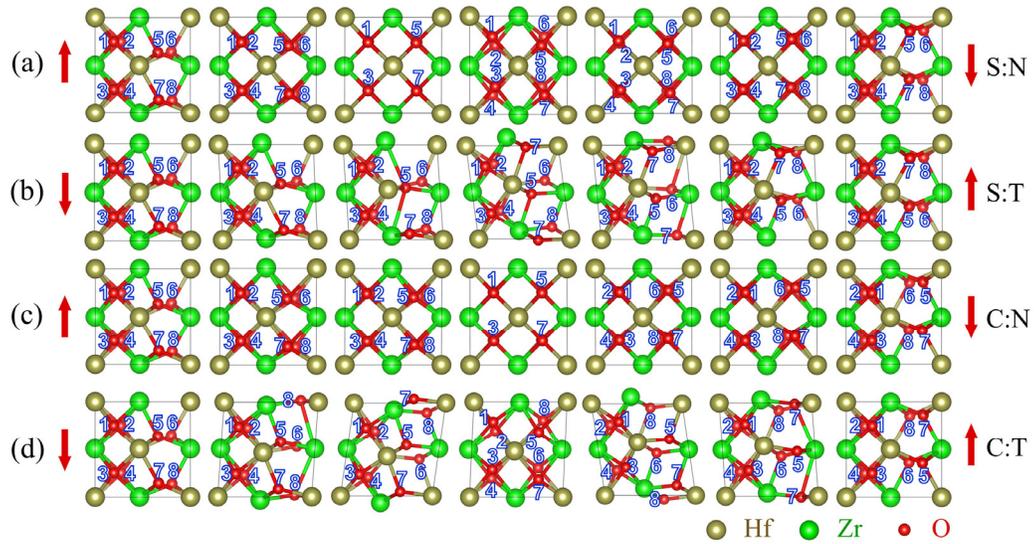

FIG. S4. The structural evolution of four uniform polarization reversal pathways of HZO calculated within VCNEB method. (a) S:N pathway. (b) S:T pathway. (c) C:N pathway. (d) C:T pathway.

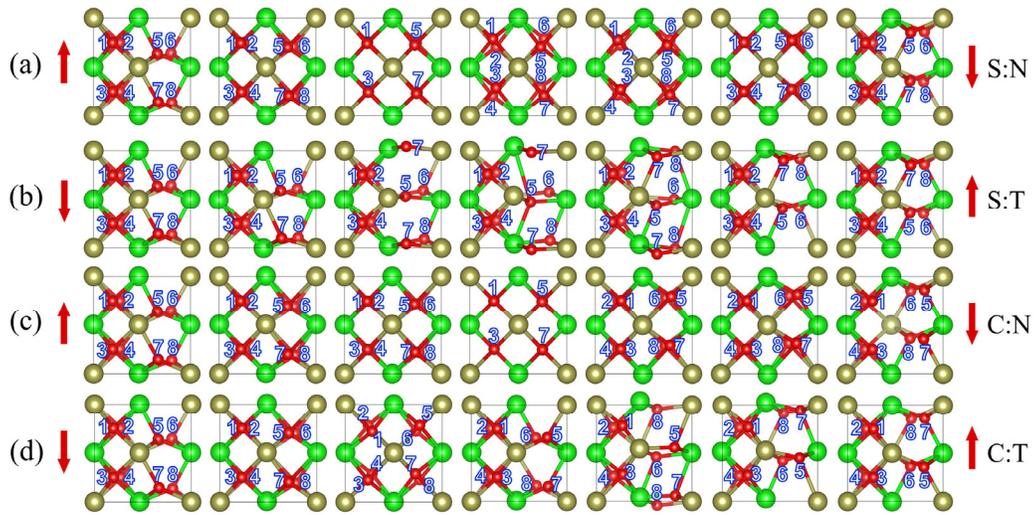

FIG. S5. The structural evolution of four uniform polarization reversal pathways of HZO calculated within nVCNEB method. (a) S:N pathway. (b) S:T pathway. (c) C:N pathway. (d) C:T pathway.

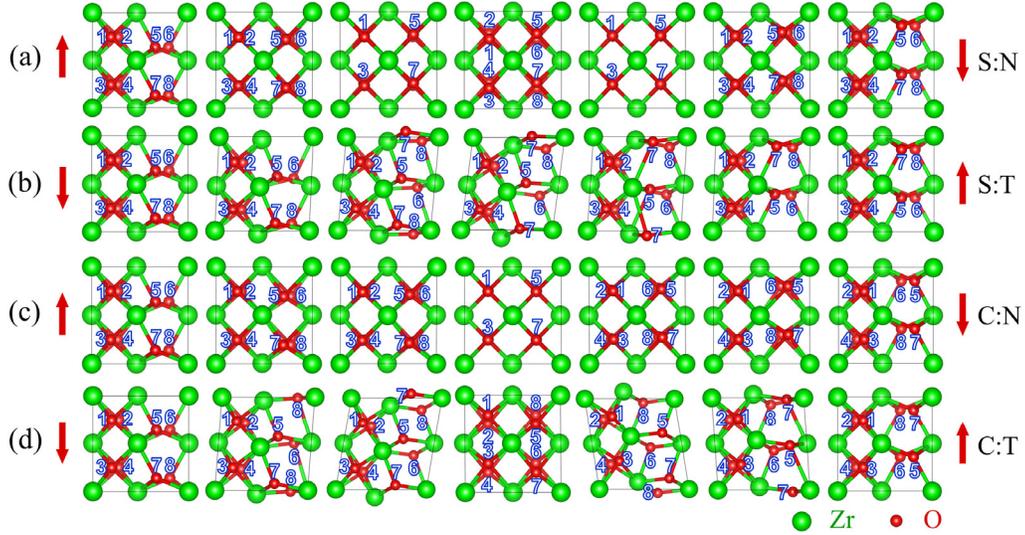

FIG. S6. The structural evolution of four uniform polarization reversal pathways of $ZrO_2$ calculated within VCNEB method. (a) S:N pathway. (b) S:T pathway. (c) C:N pathway. (d) C:T pathway.

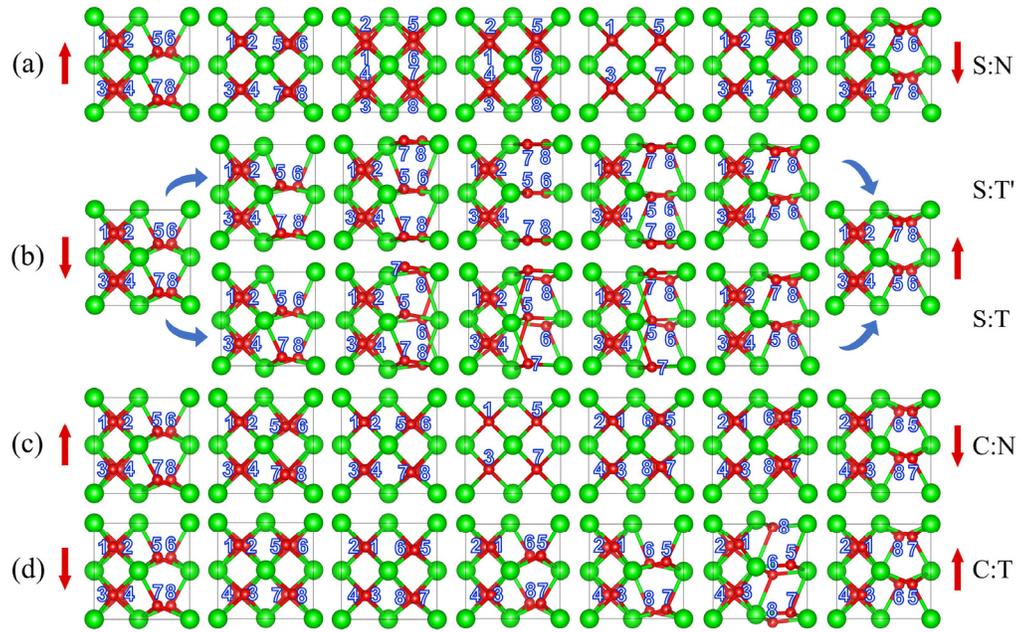

FIG. S7. The structural evolution of four uniform polarization reversal pathways of $ZrO_2$ calculated within nVCNEB method. (a) S:N pathway. (b) S:T' (top panel) and S:T (bottom panel) pathways. (c) C:N pathway. (d) C:T pathway.

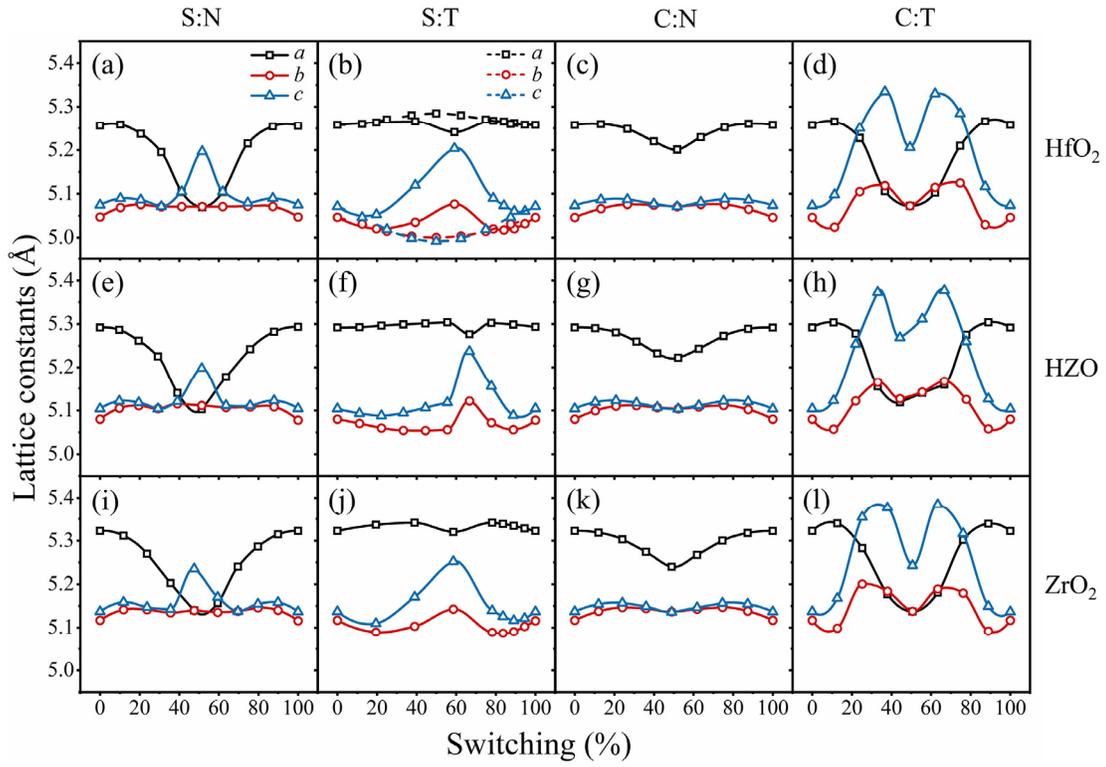

FIG. S8. The lattice constants evolution of four uniform polarization reversal pathways of $HfO_2$, HZO, and $ZrO_2$ calculated within VCNEB method. The dashed and solid lines in (b) represent the lattice constants variation of S:T' and S:T pathways, respectively.

## III. Supplementary structural evolution of DW motion

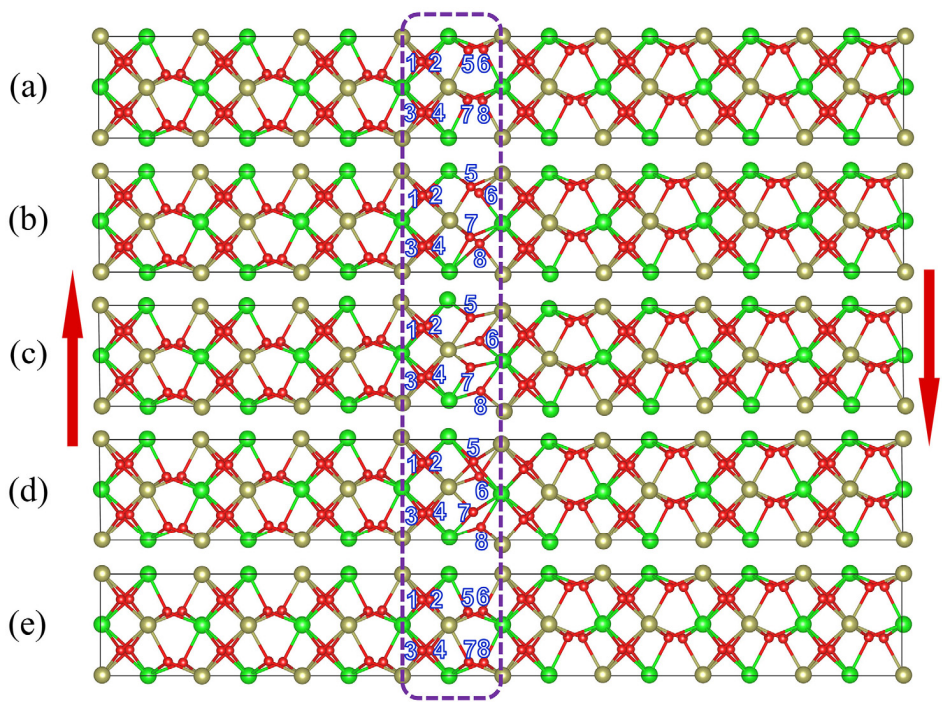

FIG. S9. The structural evolution of DW motion in HZO along S:N pathway. The rectangle denotes the unit cell with polarization reversal. Red arrows represent the directions of the ferroelectric polarizations of the two neighboring domains.

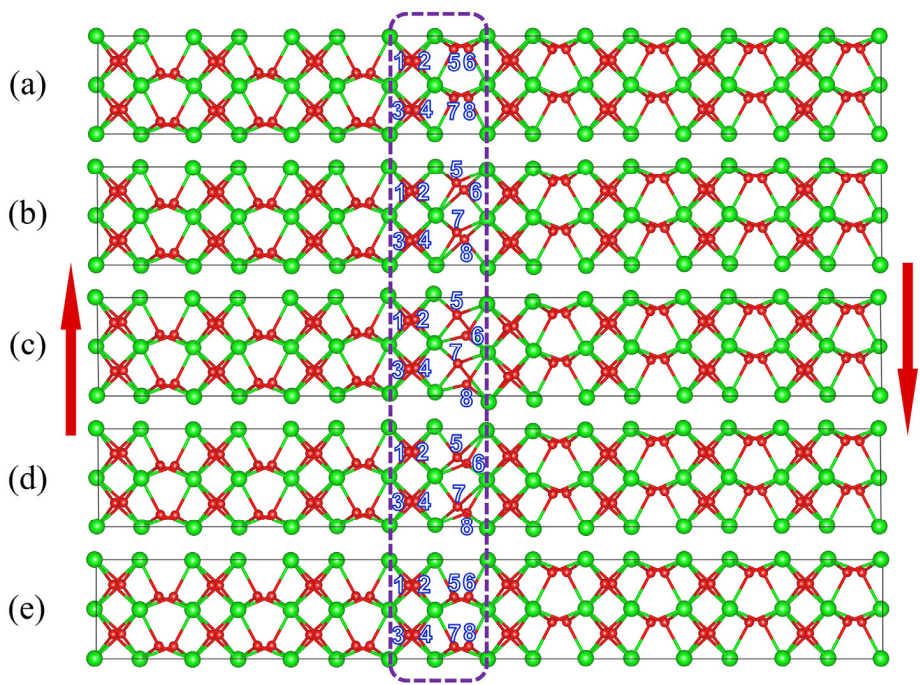

FIG. S10. The structural evolution of DW motion in $ZrO_2$ along S:N pathway.

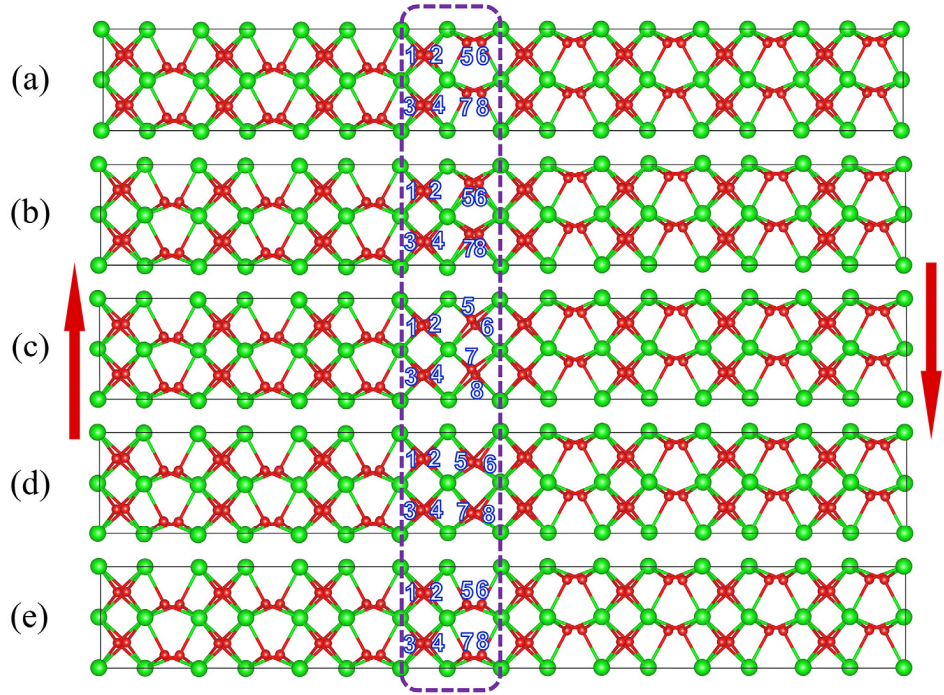

FIG. S11. The structural evolution of DW motion for ZrO$_2$ along S:N' pathway.

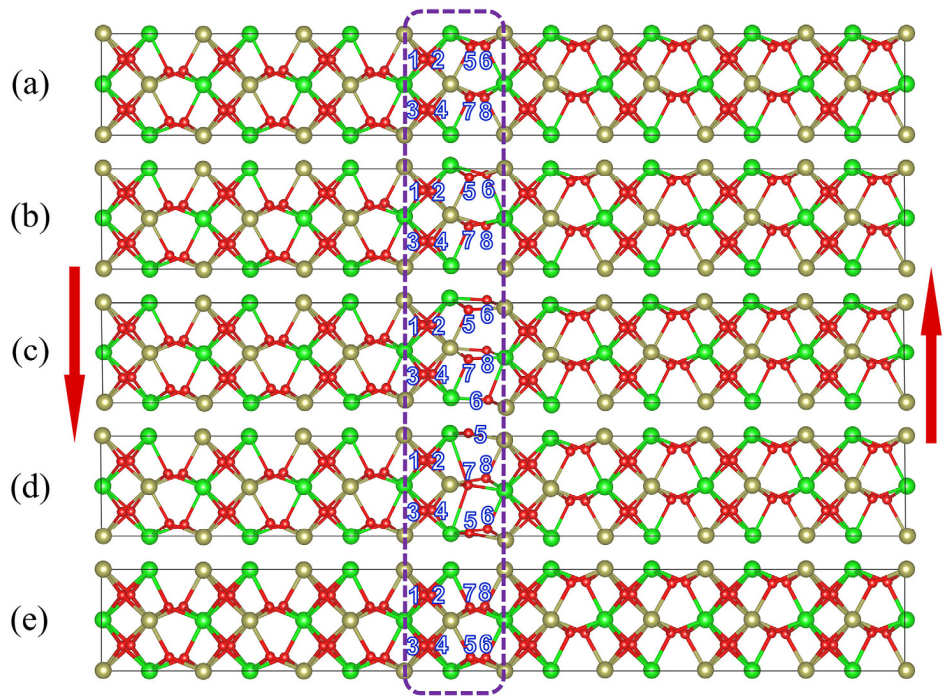

FIG. S12. The structural evolution of DW motion for HZO along S:T pathway.

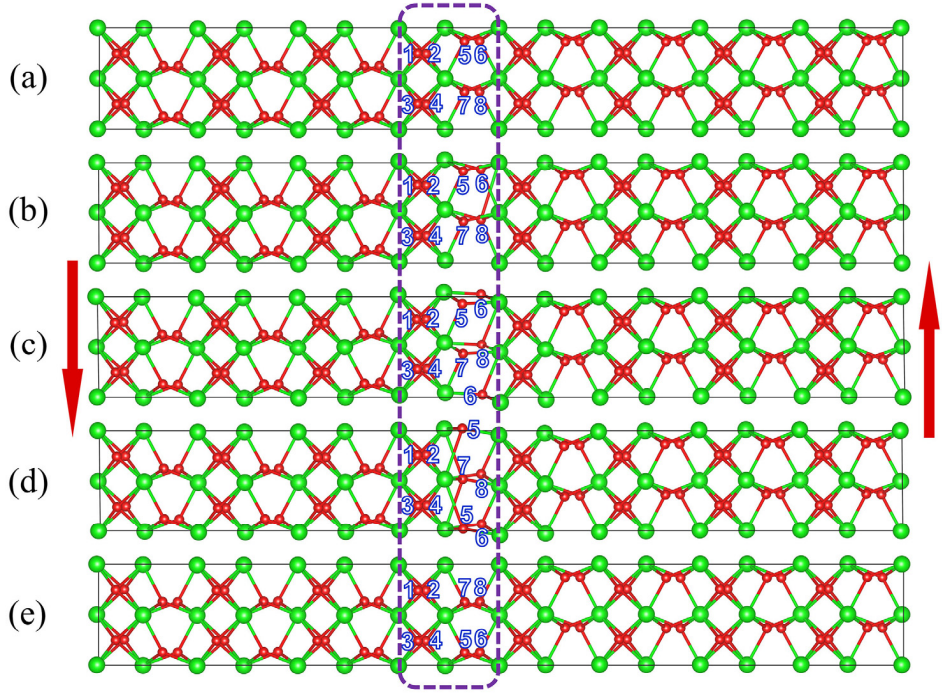

FIG. S13. The structural evolution of DW motion for ZrO$_2$ along S:T pathway.

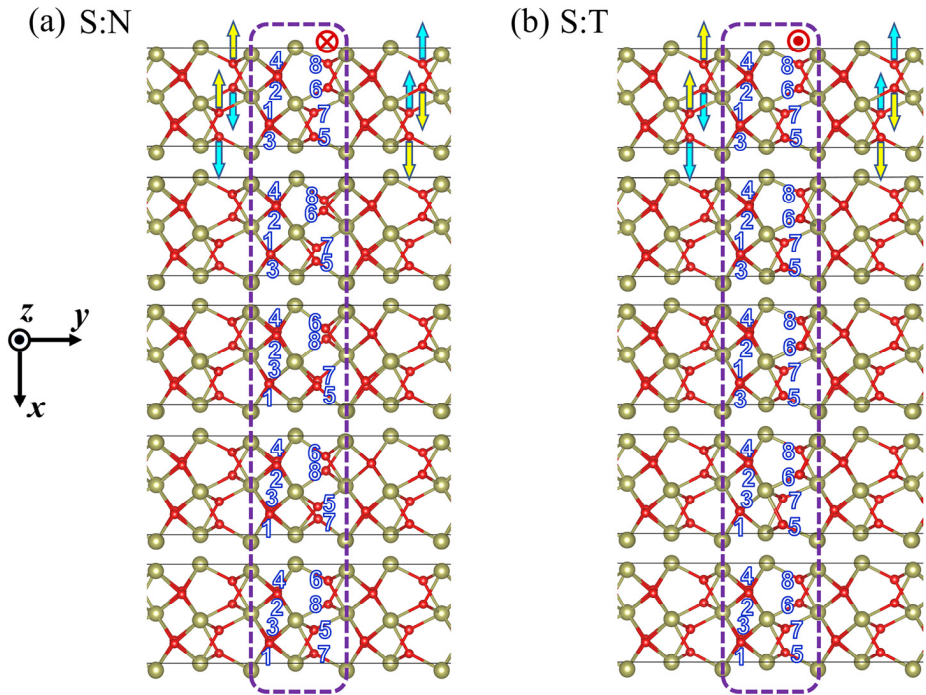

FIG. S14. The structural evolution of DW motion in HfO$_2$ along (a) S:N and (b) S:T pathways shown in the plane normal to polarization (the *z*-direction). Yellow and cyan arrows denote the opposite displacements

of top and bottom layers of three-fold coordinated O atoms along the *x*-direction, respectively. Red inward and outward symbols indicate the directions of motion of the three-fold coordinated O atoms along the polarization direction (*z*-direction) for (a) S:N and (b) S:T pathways.

**IV. One unit cell polarization reversal in a single domain (nucleation of a domain with reversed polarization)**

As sketched in Figs. S15(a) and (b), one unit cell (UC) polarization reversal in a single domain of $HfO_2$, HZO, and $ZrO_2$ ferroelectrics is studied. The energy profiles are shown in Figs. S15(c-e). For the S:N pathway, both the simultaneous (named S:N' pathway) and successive (named S:N pathway) upward shift modes of three-fold coordinated O atoms are found stable based on the simulation. The energy barriers for $HfO_2$, HZO, and $ZrO_2$ along the S:N' pathway are calculated to be 1.329, 1.229, and 1.073 eV with a single barrier (black dashed lines in Figs. S15(c-e)), respectively, which are in good agreement with the results of Lee *et al* (1.34 eV for $HfO_2$) [26]. The corresponding structural evolutions for the three ferroelectrics are plotted in Figs. S16-S18, which show that the intermediate states of the unit cell are close to the tetragonal *P*4$_2$/*nmc* phase. In contrast, the energy barriers of $HfO_2$, HZO, and $ZrO_2$ along the S:N pathway are calculated to be 0.918, 0.860, and 0.786 eV (black solid lines in Figs. S15(c-e)), much lower than those along the S:N' pathway. The corresponding structural evolution plotted in Figs. S15(f), S19, and S20 shows that the intermediate states are close to the monoclinic *P*2$_1$/*c* phase. The energy profiles of the UC polarization reversal in a single domain along the S:T pathway in $HfO_2$, HZO, and $ZrO_2$ are shown in Figs. S15(c-e) (red solid lines) with the barrier heights calculated to be only 0.517, 0.480, and 0.427 eV, respectively (nearly a half of those along the S:N pathway). From Figs. S15(g), S21, and S22, the intermediate states for the UC polarization reversal are also close to the *P*2$_1$/*c* phase due to the successive downward shift of the three-fold coordinated O atom.

nVCNEB simulations give similar energy barriers (Fig. S23 and Table SIII) and structural evolution (not shown) for both DW motion and UC polarization reversal in a single domain compared to the VCNEB simulations. This is because the local dynamic strain at the DW in the two simulation frameworks can be tolerated in a large domain configuration.

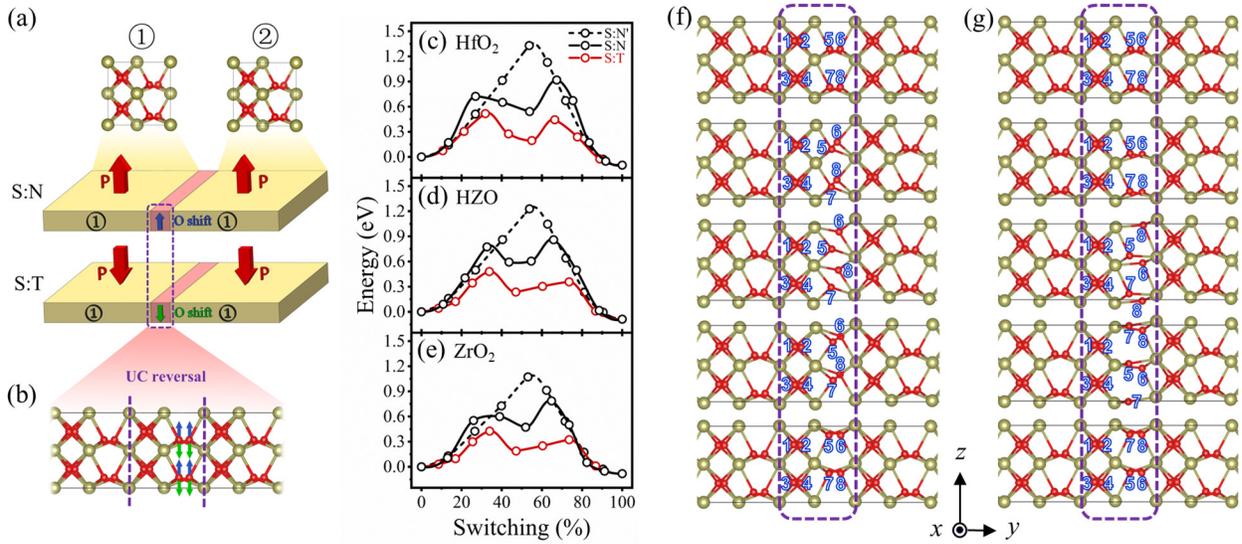

FIG. S15. (a) The sketch of UC polarization reversal in a single domain of (Hf, Zr)$O_2$ ferroelectrics along S:N and S:T pathways. Red arrows represent the directions of the ferroelectric polarizations in the single domains. (b) The shifting directions of three-fold coordinated O atoms for the UC polarization reversal along S:N (blue arrows) and S:T (green arrows) pathways. (c)-(e) The energy landscapes of UC polarization reversal in (c) $HfO_2$, (d) HZO, and (e) $ZrO_2$, respectively. The total energy of the single domain (without DW) is set to be 0. (f)-(g) The atomic evolution of UC polarization reversal in $HfO_2$ along (f) S:N and (g) S:T pathways, respectively. Rectangles denote the unit cell with polarization reversal.

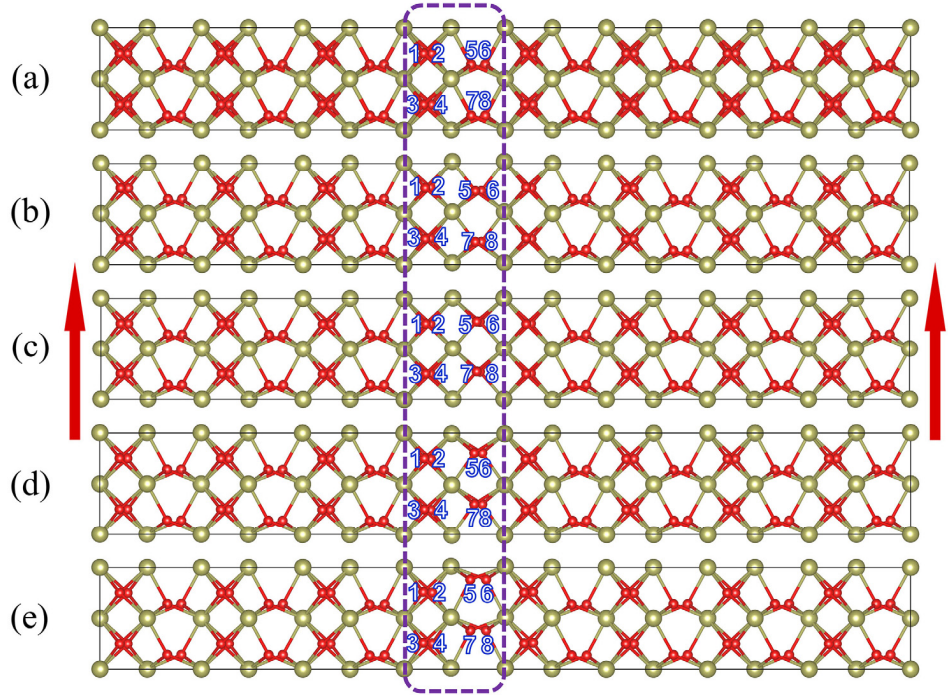

FIG. S16. Structural evolution of one unit cell polarization reversal in a single domain of $HfO_2$ along S:N' pathway. Red arrows represent the direction of the ferroelectric polarization of single domain.

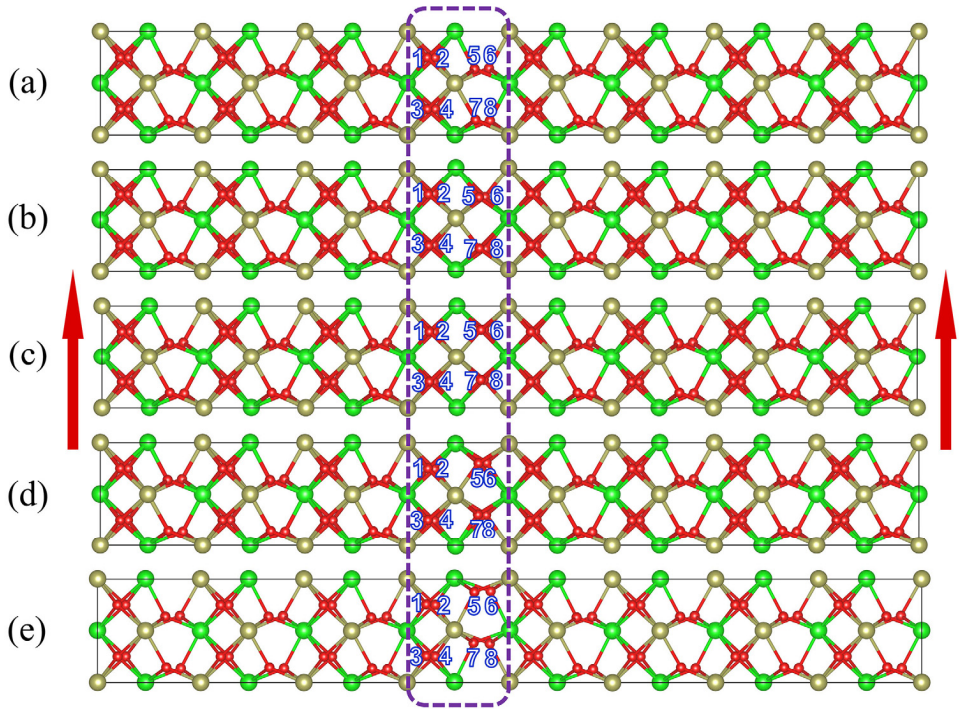

FIG. S17. The structural evolution of UC reversal in a single domain of HZO along S:N' pathway.

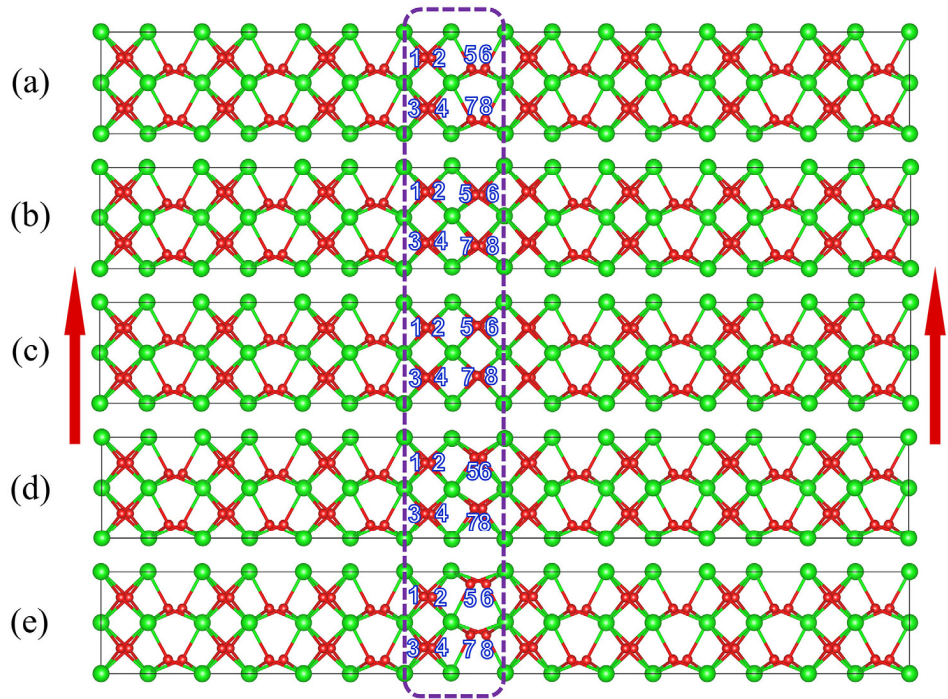

FIG. S18. The structural evolution of UC reversal in a single domain of $ZrO_2$ along S:N' pathway.

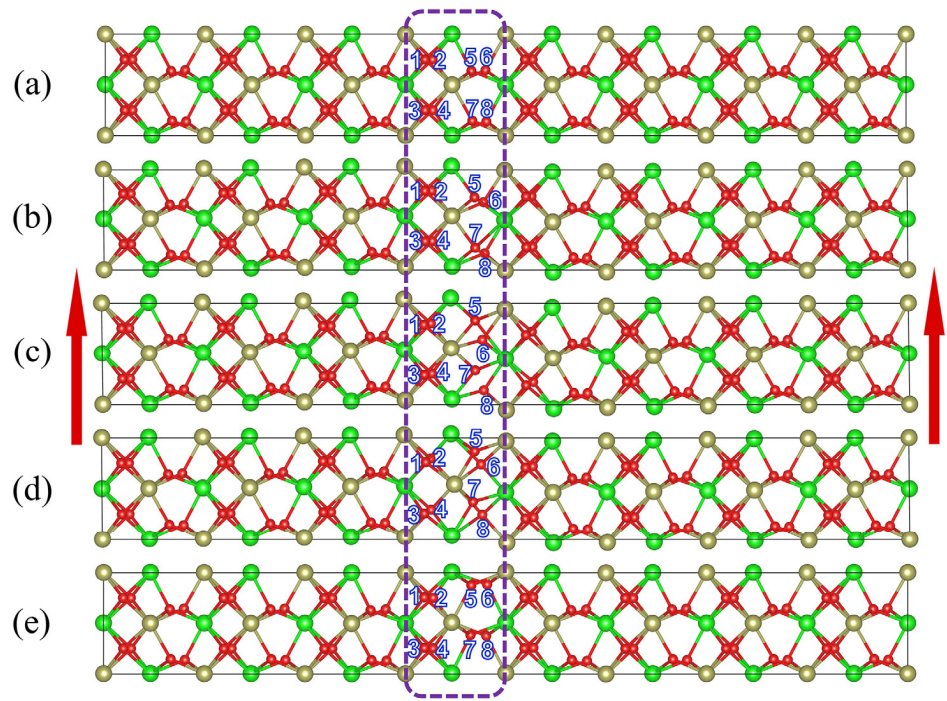

FIG. S19. The structural evolution of UC reversal in a single domain of HZO along S:N pathway.

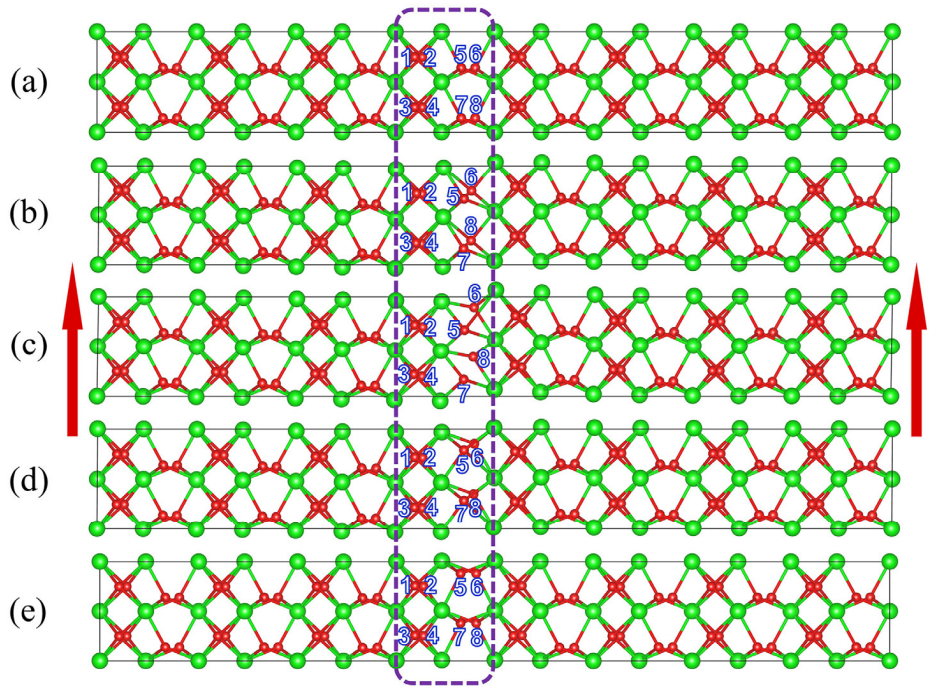

FIG. S20. The structural evolution of UC reversal in a single domain of ZrO$_2$ along S:N pathway.

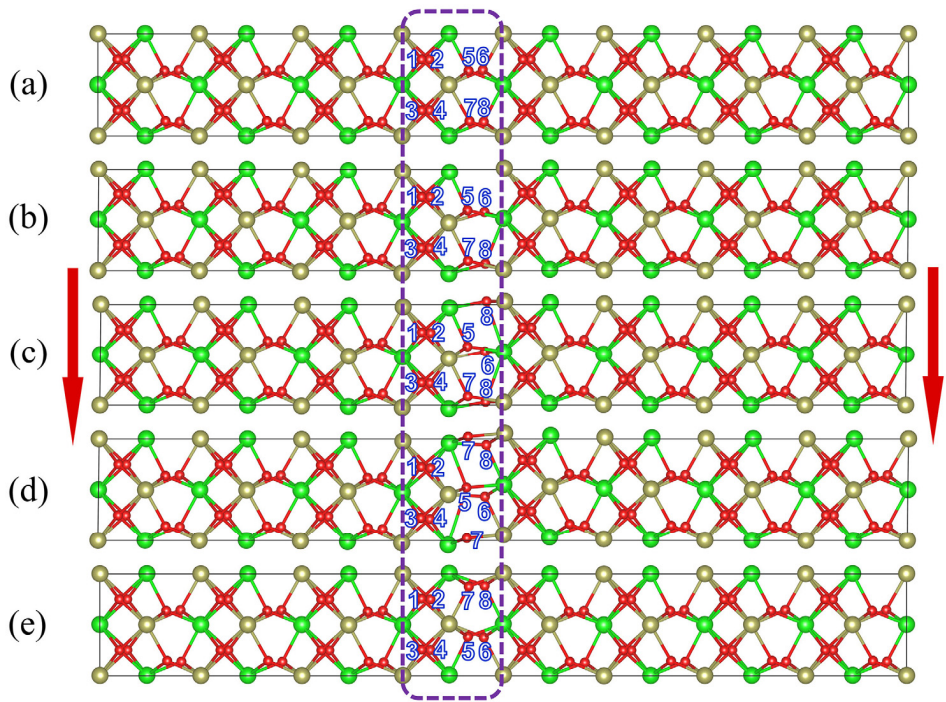

FIG. S21. The structural evolution of UC reversal in a single domain of HZO along S:T pathway.

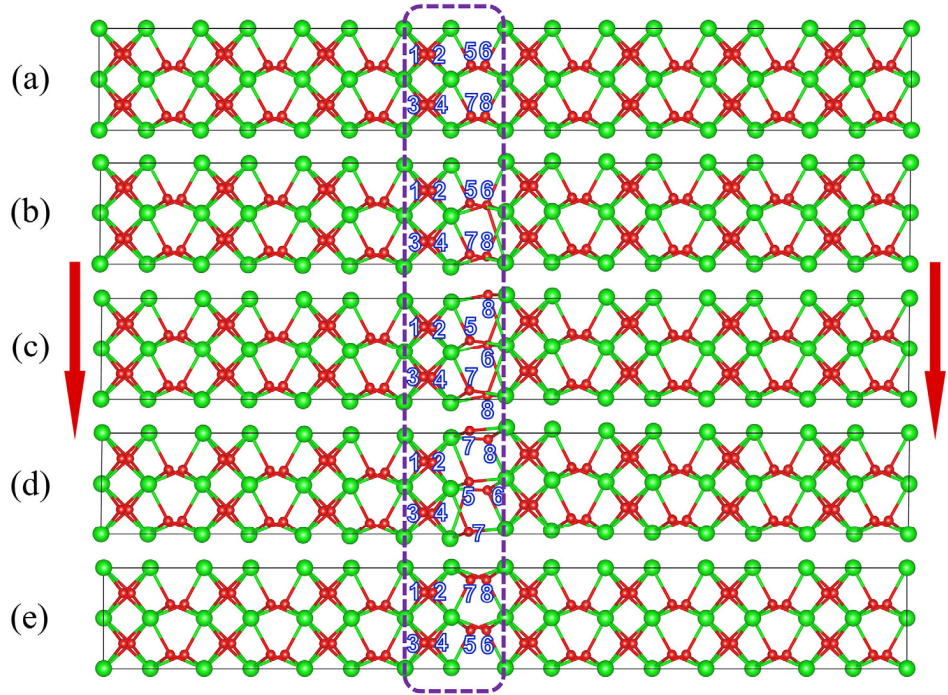

FIG. S22. The structural evolution of UC reversal in a single domain of ZrO$_2$ along S:T pathway.

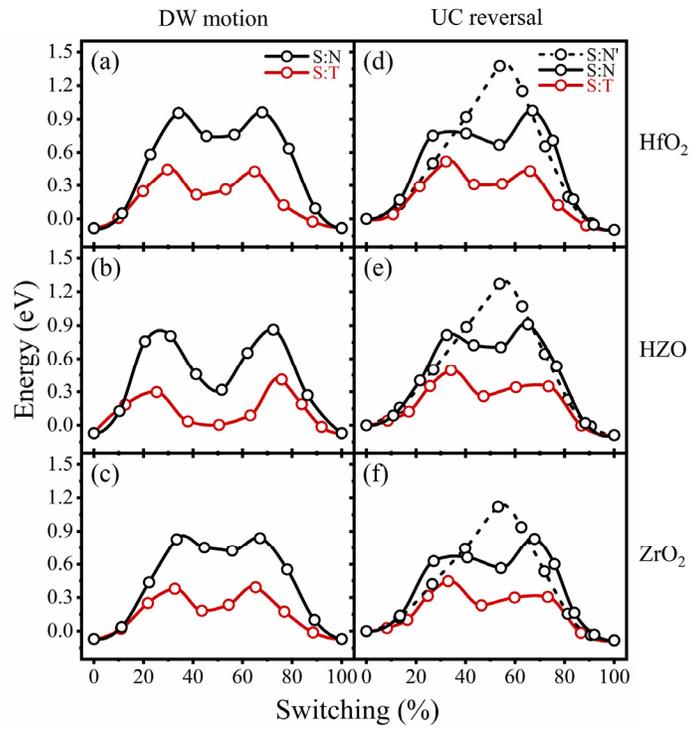

FIG. S23. The energy landscapes of (a-c) DW motion and (d-f) UC reversal in a single domain calculated within the nVCNEB method.

TABLE SIII. The energy barriers ($E_b$, per unit cell DW) of the DW motion and UC reversal in a single domain of $HfO_2$, HZO, and $ZrO_2$ calculated within VCNEB and nVCNEB methods, respectively.

| Methods | Pathways | DW motion | | | UC reversal | | |
|---|---|---|---|---|---|---|---|
| | | $HfO_2$ | HZO | $ZrO_2$ | $HfO_2$ | HZO | $ZrO_2$ |
| VC | S:N' | / | / | 0.961 | 1.329 | 1.229 | 1.073 |
| | S:N | 1.040 | 0.911 | 0.876 | 0.918 | 0.860 | 0.786 |
| | S:T | 0.509 | 0.464 | 0.452 | 0.517 | 0.480 | 0.427 |
| nVC | S:N' | / | / | / | 1.377 | 1.273 | 1.122 |
| | S:N | 1.044 | 0.935 | 0.909 | 0.976 | 0.912 | 0.832 |
| | S:T | 0.530 | 0.482 | 0.463 | 0.520 | 0.496 | 0.448 |

## VI. Spontaneous polarization based on the Berry phase method

Berry phase calculations [13] are performed to obtain the polarization of (Hf, Zr)$O_2$ along four pathways corresponding to Figs. S1(a) and (b). The results are shown in Fig. S24 and Table SI.

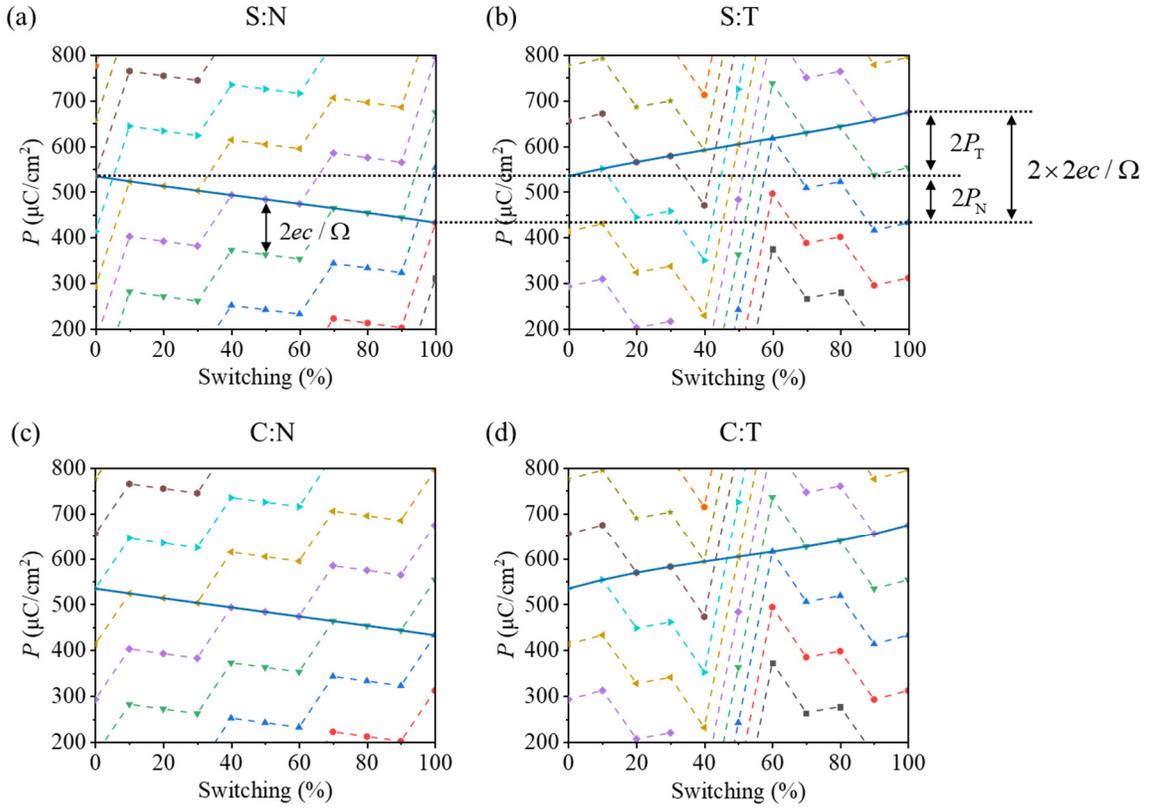

FIG. S24. The polarization evolution of $HfO_2$ along (a) S:N, (b) S:T, (c) C:N, and (d) C:T pathways. The polarization difference between the end and start points of the polarization branches (guided by the solid curves) is the double spontaneous polarization. Black arrow in (a) indicates the polarization quantum $2ec/\Omega$. Arrows in (b) indicate values of $2P_T$ and $2P_N$ with their absolute values adding up to $2\times 2ec/\Omega$.

## VII. Spontaneous polarization based on the Born effective charge method

Here we perform calculations of polarization in HfO$_2$ based on Born effective charges. We note that contrary to the conventional perovskite ferroelectrics, such as BaTiO$_3$, where the cubic *Pm-3m* phase can be used as the reference centrosymmetric structure to calculate Born charges, in HfO$_2$ of the orthorhombic *Pca*2$_1$ phase, there are two possible centrosymmetric structures: tetragonal *P*4$_2$/*nmc* and orthorhombic *Pbcm*, which could be used as a reference. The former is arguably a better representation for the N switching pathway and the latter is a better representation for the T switching pathway. In addition, the concept of Born effective charges is based on a linear relationship between the polarization change and atomic displacements. While this approximation works well for small displacements ~0.1 Å typical for perovskite ferroelectrics, the displacements of O atoms in ferroelectric HfO$_2$ between two oppositely polarized states are large (~1 Å). This makes the linear relationship between polarization and atomic displacements not accurate and leads to overestimated values of the calculated polarization. Nevertheless, as expected, the calculated polarization values based on Born effective charges are different for the N and T paths and are in qualitative agreement with those obtained using the Berry phase method.

The Born effective charge tensor $\mathbf{Z}_n^*$ for Hf and O atoms in HfO$_2$ is calculated based on the standard procedure [27] using $\Delta \mathbf{P} = \frac{e}{\Omega} \sum_n \mathbf{Z}_n^* \Delta \mathbf{u}_n$, where $\Delta \mathbf{P}$ is a change in polarization $\mathbf{P}$ under displacement $\Delta \mathbf{u}_n$ of atom $n$ and $\Omega$ is the unit cell volume. The results for the diagonal elements of the Born effective charge tensor for Hf and 3-fold coordinated O atoms are presented in Table SIV for *P*4$_2$/*nmc* and *Pbcm* structural phases as a reference.

TABLE SIV. The diagonal elements of the Born effective charge tensor $\mathbf{Z}_n^*$ in HfO$_2$ calculated for Hf and 3-fold coordinated O atoms for the *P*4$_2$/*nmc* and *Pbcm* reference phases.

| Phase for Born charges | $\mathbf{Z}_n^*$ component | Hf | O |
|---|---|---|---|
| *P*4$_2$/*nmc* | xx | 5.54 | –2.77 |
| | yy | 5.54 | –2.77 |
| | zz | 4.84 | –2.42 |
| *Pbcm* | xx | 5.77 | –2.80 |
| | yy | 5.63 | –3.00 |
| | zz | 4.28 | –1.99 |

Using these Born effective charges, we then calculated the polarization values *P* by displacing the HfO$_2$ atoms along the S:N and S:T pathways between oppositely polarized states, which are allowed for the domain-wall induced switching. Table SV shows the obtained polarization values.

TABLE SV. Polarization magnitudes (in units of µC/cm²) calculated based on the Born effective charge method in comparison to those calculated within the Berry phase method for the S:N and S:T pathways.

| Pathway | Reference phase | Phase for Born charges | Polarization: Born charge method | Polarization: Berry phase method |
|---|---|---|---|---|
| S:N | $P4_2/nmc$ | $P4_2/nmc$ | 62.5 | 51.0 |
|  |  | $Pbcm$ | 52.3 |  |
| S:T | $Pbcm$ | $P4_2/nmc$ | 83.6 | 69.7 |
|  |  | $Pbcm$ | 70.0 |  |

We see that the polarization values obtained from the Born charges based on both reference phases are in a qualitative agreement with those calculated using the Berry phase method. Importantly, these values are clearly different for the S:N and S:T pathways reflecting the different direction of ionic motion during the polarization switching process.

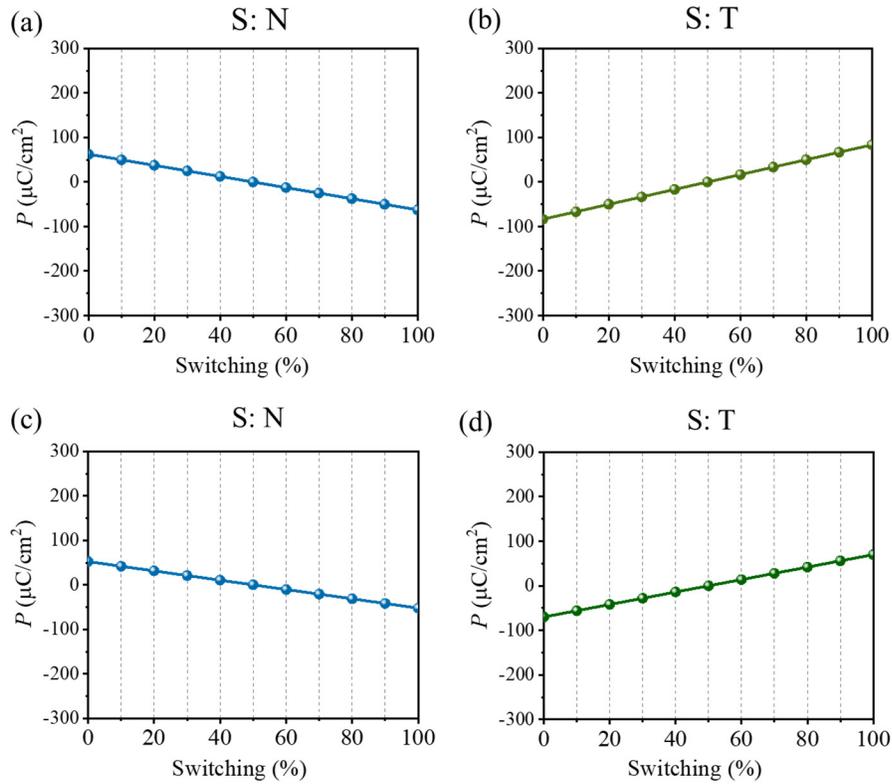

FIG. S25. Polarization evolution for the S:N and S:T pathways obtained within the Born charge method using effective charges calculated from the $P4_2/nmc$ (a,b) and $Pbcm$ (c,d) reference phases.

Fig. S25 shows the polarization evolution when switching between two oppositely polarized states in ferroelectric $HfO_2$ for the S:N and S:T pathways, as calculated within the Born charge method.

Evidently, these curves are similar to those shown in Fig. S24, demonstrating qualitative agreement between the Born charge method and the Berry phase method for polarization calculation. In particular, it is seen that the opposite slope of the curves indicate opposite polarization orientation for the S:N and S:T pathways. All these calculations clearly indicate that the polarization value is dependent on the switching pathway and is different (both in magnitude and orientation) for the S:N and S:T pathways.